\documentclass[preprint,showpacs,preprintnumbers,amsmath,amssymb]{revtex4}

\usepackage{graphicx}% Include figure files
\usepackage{dcolumn}% Align table columns on decimal point
\usepackage{bm}% bold math

\begin{document}

%\preprint{APS/123-QED}

\title{Diffusivity and Weak clustering in a Quasi 2D Granular Gas}

\author{J.~A.~Perera-Burgos$^{1,2}$}
\email{jperera@mda.cinvestav.mx}
%\altaffiliation[Also at ]{Physics Department, XYZ University.}
\author{G.~P\'erez-\'Angel$^{1,2}$}
\author{Y.~Nahmad-Molinari$^2$}
\affiliation{$^1$Departamento de F\'{\i}sica Aplicada, Centro de Investigaci\'on y de
Estudios Avanzados\\
del IPN, Unidad M\'erida, AP 73 ``Cordemex'', 97310 M\'erida, Yuc. 
M\'exico;}
\affiliation{$^2$Instituto de F\'isica ``Manuel Sandoval Vallarta'', Universidad Aut\'onoma 
de San Luis Potos\'i, \'Alvaro Obreg\'on 64, San Luis Potos\'i,
SLP, M\'exico}

\date{\today}

\begin{abstract}
We present results from a detailed simulation of a 
quasi-2D dissipative granular gas, kept in a non-condensed steady state via 
vertical shaking over a rough substrate. This gas shows a weak power-law decay in 
the tails of its Pair Distribution Functions (PDF's), indicating fractality and 
therefore a tendency to form clusters over several size scales. This clustering 
depends monotonically on the dissipation coefficient, and disappears when
the sphere-sphere collisions are conservative. 
Clustering is also sensitive to the packing fraction.
%Effective potentials for the interparticle interaction are 
%extracted for every PDF using the Ornstein-Zernike equation with the Percus-Yevick 
%closure. 
This gas also displays the standard non-equilibrium 
characteristics of similar systems, including non-Maxwellian velocity 
distributions. The diffusion coefficients are calculated over all the conditions of 
the simulations, and it is found that diluted gases are more diffusive
for smaller restitution coefficients.
\end{abstract}

\maketitle

\section{Introduction} 
Two--dimensional (2D) granular gases have been widely studied as 
examples of out-of-equilibrium systems that are both simple enough to analyze and 
easy to construct experimentally. In the pioneering work of Olafsen and Urbach 
\cite{Urbach_PRL81,Urbach_PRE60} it was found that these gases, kept in a steady 
state via vertical shaking, could condensate into an hexagonal solid phase (for 
monodisperse sample) as the amplitude or frequency of shaking is reduced. This gives 
a fascinating example of fluid-solid transition not driven by molecular attraction 
or entropy maximization. In the gaseous phase, and confining the vertical
expansion of the 
granular layer, the 2D
Velocity Distribution Functions (VDFs) displayed by this system are 
non-Maxwellian, and tend to fall instead into an stretched exponential form. The 
deviation from Maxwellian behavior of the VDFs in granular gases 
has been studied theoretically 
\cite{Puglisi_PRE59,Brilliantov_PRE61,van_Noije_Granular_Matter2008}, in simulations 
\cite{Taguchi_EPL30,Nie_EPL51,Barrat_PRE66,MacKintosh_PRL93,MacKintosh_PRE72,Bray_PRE75,Burdeau_PRE79}, 
and in experiments 
\cite{Urbach_PRE60,Losert_Chaos09,Kudrolli_PRE62,Prevost_PRL89,Olafsen_PRL95,Reis_PRE75}, 
and have been characterized using Sonine polynomials 
\cite{van_Noije_Granular_Matter2008,Reis_PRE75}. 
%Nevertheless it is necessary to be 
%careful when the expansion is done, because there is a breakdown of the higher-order 
%Sonine coefficients caused by the increasing impact of the overpopulated high-energy 
%tail of the distribution function \cite{Brilliantov_EPL74}.

In its simpler form, experiments in granular gases are carried on using a horizontal 
cell with a perfectly flat bottom; due to dissipation, however, horizontal components 
of momentum tend to decrease an thus the gas finally undergoes an inelastic collapse 
into a static (with respect to the plane) condensed phase. In shaken experiments,
for large vertical 
acceleration, and in the absence of vertical confinement, there is some dispersion in the 
instantaneous vertical position of the spheres, and the resulting off-plane 
collisions serve as a way of converting vertical momentum into horizontal momentum, 
keeping in this way the gas from condensing. This effect is large enough that it 
allows for the gas to keep some horizontal motion even in cases 
where a fraction of the beads 
never loose touch with the bottom \cite{Urbach_PRL81}. 
In recent years some attempts have been made to 
provide mechanisms by which this transfer from vertical to horizontal momentum can 
be effected without depending on the fluctuations in height of the beads. For 
instance many theoretical and numerical models have been proposed based on an 
homogeneous granular gas randomly driven by a white noise %energy source 
\cite{van_Noije_Granular_Matter2008,Williams_and_MacKintosh_PRE54}. Experimentally, 
a complete layer of spheres glued to the bottom plate (the ``floor''), or a first layer of heavy 
spheres on which a layer of light spheres is placed, have been used 
\cite{Prevost_PRL89,Baxter_Nature2003}, and in other cases an artificially roughened 
bottom plate was employed \cite{Reis_PRE75,Ruth_EPJE28}.
It was found in these experiments that the behavior of the gas was somewhat 
closer to that of an ideal hard disk gas, 
including velocity distributions closer to the Maxwellian. The use of a roughened 
substrate makes the movement of the particles to resemble Brownian motion, because 
of the frequent scatterings with the uneven floor.

Motivated by the experimental results reported in \cite{Ruth_EPJE28}, 
in this paper we are describing a fully three-dimensional simulation
of a confined vertically-shaken granular gas, where structure has been 
imposed in the bottom 
plate using fixed and non-overlapping small hemispheres. We find that this mechanism is
quite efficient in avoiding the collapse of the gas into the quiescent state (that
is, a state with no horizontal motion).
We record the Mean Square Displacement (MSD) of the spheres in the gas, and study the dependence
of the diffusion coefficient on packing fraction, adimensional acceleration and on restitution coefficient.
We also report the obtained horizontal velocity distributions, and show the strong effects
that friction with the upper confining plate (the ``ceiling'') have in these functions.
Finally we characterize the instantaneous state 
of the gas using its Pair Distribution Functions (PDF's) 
\cite{Urbach_PRL81,Urbach_PRE60,Tata_PRL84,P.Eshuis_PRL95,Pacheco_PRL102}, 
which give us information about the underlying effective interparticle interactions. 
We observe evidence of week clustering in these PDFs, and 
relate them to the changes in diffusion in the system.
%use them, together with the Orstein-Zernike equation with the Perkus-Yevick closure,
%to extract effective potentials for the gas. 

\section{Simulation} 

\begin{figure} %\onefigure[width=6.5cm]{figuras/gas_granular.ps} 
\includegraphics[width=12.0cm,angle=0]{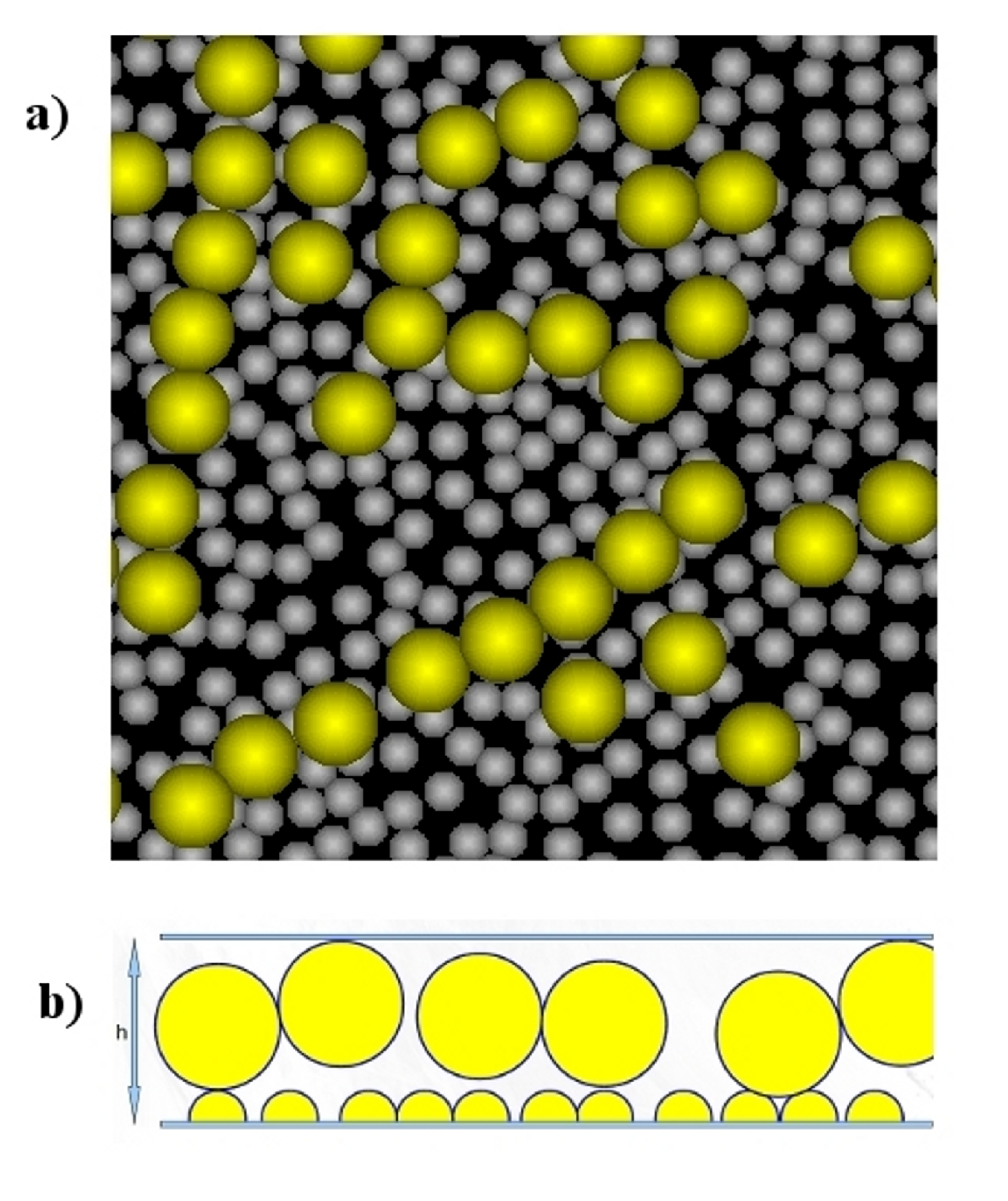} 

\caption{(a) Fraction of a snapshot (taken from above) of the granular gas, covering around $1/13$
of the actually simulated area. The hemispheres (scatterers)
are shown as white circles, the moving grains as yellow circles.
Notice the presence of some short linear chains.
(b) Schematic lateral view of the simulated system, to scale.} 
\label{top_view_gg} 
\end{figure}

We perform numerical simulations of a monodisperse granular 
gas of spheres in a quasi-bidimensional space, square with side $L$
and with periodic 
boundary conditions in the horizontal. This gas is confined between two horizontal 
planes which oscillate in the vertical in a sinusoidal way $z = A\, \sin 
(\omega\, t)$, with amplitude $A$ and frequency $\omega$ large enough so that 
the spheres will in general touch both confining plates in every 
oscillation of the system.
In the bottom plate there are fixed and non-overlapping
hemispheres whose diameter $\sigma_{hs}$ is almost half 
of the diameter of the free spheres $\sigma_{s}$. 
These hemispheres move synchronously with the plane and are randomly 
distributed with a sufficiently high two-dimensional packing fraction 
$\phi_{hs}$, so that no wide flat patches can be formed
(see Fig.~(\ref{top_view_gg})). 
Here $\phi_{hs} = N_{hs} \pi \sigma_{hs}^2/(4 L^2)$, where $N_{hs}$ is the 
number of hemispheres in the plate of side $L$.
The separation between planes is 
$h = 1.6 \sigma_{s} + \sigma_{hs}/4 \approx 1.7 \sigma_s$,
and the minimum distance between the centers of two hemispheres is $d_m$.

In order to solve the Newton's equations of motion we use standard time-driven 
molecular dynamics, using a velocity-Verlet integration with a predictor step for
velocities. This is necessary in order to incorporate the dissipative part of the force, 
which is velocity-dependent. The force between particles in contact
is described by 
    \begin{equation} 
    {\bf F}_{ij} =
    \left\{
    \begin{array}{cl}
    {\bf F}_{ij}^n + {\bf F}_{ij}^t & \mbox{when $\xi_{ij} > 0$}, \\
    0 & \mbox{otherwise},
    \end{array}
    \right. 
    \end{equation} 
where ${\bf F}_{ij}^n$ is a normal force which causes 
changes of the translational motion of the particles and ${\bf F}_{ij}^t$ is a 
tangential force, originating in friction,
which causes changes in both the translational and the rotational 
motion. The quantity $\xi_{ij}$ is the mutual overlap (compression)
 of particles $i$ and $j$ 
and is defined by
    \begin{equation}
    \xi_{ij} = \mbox{max}(0, \sigma_{s} - |{\bf r}_{i} - {\bf r}_{j}|) 
    \end{equation} 
for interparticle collisions, by
    \begin{equation}
    \xi_{ij} = \mbox{max}(0, (\sigma_{s} + \sigma_{hs})/2  - |{\bf r}_{i} - {\bf R}_{j}|) 
    \end{equation} 
for particle-scatterer collisions, and by
    \begin{equation}
    \xi_{ip} = \mbox{max}(0, \sigma_{s}/2 - d_{ip}) 
    \end{equation} 
for sphere-plate collisions. Here ${\bf r}_i$ denotes the center of 
the $i$-th grain, ${\bf R}_i$ the center of the $i$-th scatterer, and
$d_{ip}$ is the normal distance from the center of the $i$-th grain to 
the surface of the plate $p$.

Modeling a force that leads to inelastic collisions requires at least two terms: 
repulsion and some type of dissipation. The existing models and their 
characteristics, as well as comparisons among them, can be found in 
\cite{Schafer_JPhys6,Brilliantov_PRE53,Di_Renzo_ChES59,Stevens_PT154,Kruggel_PT171,Book_Poschel}. 
Here we have used for the normal force the linear spring-dashpot model, in 
which the contact interaction is modeled by the damped harmonic oscillator force 
    \begin{equation}
    F_n = \mbox{min} \left\{0, \, -\kappa_n \xi -\gamma_n \dot \xi \right\}, 
    \label{eq_normal_force}
    \end{equation}
where 
$\gamma_n$ is a damping constant and $\kappa_n$ is related to the stiffness of a 
spring whose elongations is $\xi$, the overlap between two grains (or a grain and
a boundary plate). The advantage of this model lays in its analytic solution, 
where the collision time is given by
    \begin{equation} 
    t_n = \pi\,\left( \frac{\kappa_n}{m_{\mathrm{eff}}} - 
    \left( \frac{\gamma_n}{2\,m_{\mathrm{eff}}}\right)^2 \right)^{-1/2},
    \label{eq_time_coll}
    \end{equation} 
and the restitution coefficient is 
    \begin{equation} 
    e_n = \exp \left( -\frac{\gamma_n}
    {2\, m_{\mathrm{eff}}}\,t_n\right).
    \label{eq_restitut_coeff}
    \end{equation} 
Here $m_{\mathrm{eff}}$ is the effective mass for the colliding pair.
For the tangential force we consider the Coulomb dynamical friction law 
${\bf F}_{ij}^t = -\mu F_{ij}^n {\bf \hat v}_{i,j}^t$, where ${\bf v}_{i,j}^t$ 
is the unitary vector in the direction of the 
relative tangential velocity between spheres $i$ and $j$ (or, by extension, between a 
sphere and a scatterer or a sphere and a plate). The effect of 
gravity is incorporated into the vertical acceleration.

For the actual simulations we have taken most parameters similar to those of the
previously mentioned experiments \cite{Ruth_EPJE28}, as performed with steel
spheres. Therefore, we have fixed the frequency of oscillation of the cell 
to $f = 60$ Hz, and have considered two amplitudes of oscillation: $A_1 = 0.024$ cm and 
$A_2 = 0.05$ cm, for every packing fraction $\phi$ of free spheres, 
getting in this way for the adimensional acceleration 
$\Gamma \equiv (2 \pi f)^2 A/g$ the values  $\Gamma_{1} = 3.5$ 
and $\Gamma_{2} = 7.2$. Here $\phi$ is defined as before 
by $\phi = N \sigma_s^2 \pi / (4 L^2)$.
As parameters for the normal force, Eq.~(\ref{eq_normal_force}), we use
those extracted from Eqs.~(\ref{eq_time_coll},\ref{eq_restitut_coeff}), and fix
the collision time to $t_n = 7.1\mbox{x}10^{-5}$ s, and use 
three values of the restitution coefficients
$e_n = 0.36, 0.66$ and $1$ for the sphere-sphere and sphere-hemisphere interactions
(at each value of $\Gamma$ and for every packing fraction $\phi$ considered).
The coefficient of restitution for sphere-substrate interaction was fixed at $0.878$,
corresponding to an experimentally measured steel-acrylic 
restitution coefficient \cite{Ruth_EPJE28}, and the considered coefficient of 
dynamical friction $\mu$ was $0.25$.
The time-step of the integration is always fixed to 
$t_n/100$, and the simulation time is $40$ s, after a transient of $10$ s. 
The diameter of the free spheres 
$\sigma_s$ and of the hemispheres $\sigma_{hs}$ have been fixed to 
$0.44$ cm and $0.2$ cm, respectively, in all simulations. The minimal 
distance between hemispheres is 
$d_m = 0.2$ cm. Finally, the boundary plates have side $L = 16$ cm, and 
the packing fractions used have been $\phi = 0.15$, $0.20$, $0.25$, $0.30$ and $0.35$ for both $\Gamma$ 
values used. The resulting numbers of grains are
$N_s = 252, 336, 420, 504$ and $588$ 
% for $\phi = 0.15, 0.20, 0.25, 0.30$ and $0.35$
respectively, and the number of scatterers is $N_{hs} = 4100$.
All the quantities reported here are the average over $10$ simulations, each with a different 
configuration of hemispheres and different initial conditions. 

\section{Numerical result: Diffusion} 

\begin{figure} %\begin{indented} %\item[] 
\includegraphics[width=12.0cm,angle=270]{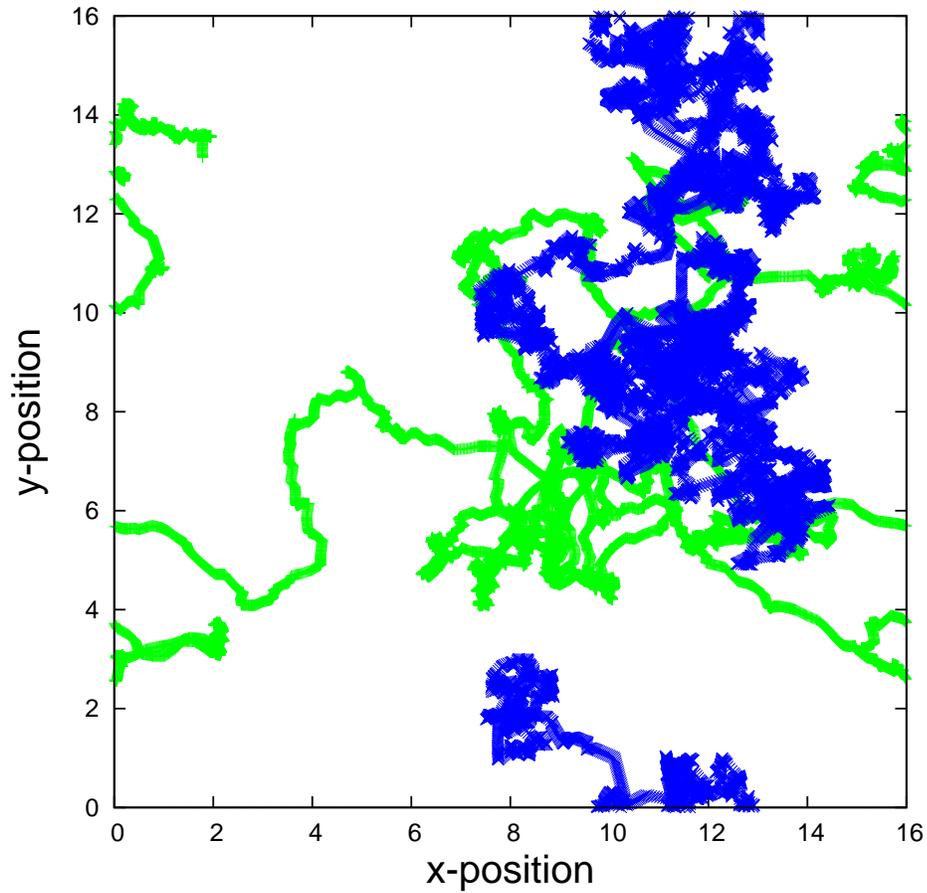} 

\caption{Trajectory of a single particle for two different restitution coefficients.
The more compact trace (blue) is for $e_n = 1.0$, the less compact one 
(green) is for $e_n = 0.36$. The packing fraction
is $0.15$ and we used $\Gamma = 3.5$. Notice how the tracer describes a much larger excursion 
for the smaller restitution coefficient.} 
\label{trayectorias} 
\end{figure}

We will start by presenting results for the diffusion in the quasi-2D granular gas.
As mentioned before, the interaction with the hemispheres provides horizontal 
momentum to the particles, but also scatters their motion, making
them to move on a pseudorandom way, resembling in certain form a 
Brownian motion. It is clear that the interparticle collisions also
contribute to the scattering. In Fig.~(\ref{trayectorias}) we can see two 
trajectories for a 
single particle on a complete simulation of 40 s, 
one for the case with $e_n = 1$ and another for $e_n = 0.36$. 
This was done with a packing fraction $\phi = 0.15$, and the trajectory
includes the interactions with the other particles 
and with the substrate. One must remember here that the restitution coefficient
for particle-plate collisions is always fixed, while that for
particle-hemisphere collisions is the same as the one for particle-particle
collisions, and can be varied.

The MSDs measured fit very well the Einstein form
   \begin{equation}
   \langle ({\bf r}(t + t_0)- {\bf r}(t_0))^2 \rangle = 4\,D\,t; 
   \end{equation}
this is not as trivially expected 
as it may seem, 
since some recent results point to a breakdown of Einstein's law for 2D
granular gases \cite{Henrique}. This is related to a well-known anomaly
for self-diffusion in 2D, where the presence of hydrodynamical backflows 
gives origin to long tails in the velocity autocorrelation function.
These tails behave as $t^{-1}$, giving in this way a logarithmic divergence
to the diffusion coefficient, accordingly to the Green-Kubo formula 
\cite{Alder_PRA1,Camp_PRE71}. In our simulations we have not seen any 
evidence of non-Einstenian behavior; in particular we have performed
sets of longer runs (80 s) for three different
sets of parameters, and one set of runs 
for a bigger system ($L = 16 \sqrt{2} = 22.63$ cm),
and in all cases the same diffusive behavior was found, with the 
diffusion coefficient independent of the length of the run or the system size.
Looking at the difference between these results and those of \cite{Henrique},
it is clear that the presence of fixed (with respect to the plane) scatterers
is the reason why regular diffusion is restored.

In Fig.~(\ref{diff}) we show 
the diffusion coefficient $D$ versus the packing
fraction for both values of the adimensional acceleration. Here we can observe the 
following points: first, $D$ increases for 
low values of $\phi$, and decrease for high values of $\phi$. These results are in 
agreement with those obtained by \cite{Zippelius,DurianPRE74}, and are to
be expected, since larger densities represent in general more obstacles to
the motion. Second, for small values of $\phi$ the diffusion coefficient increases 
when we lower $e_n$, and this behavior is stronger for 
low $\Gamma$. As the concentration and $\Gamma$ grow, this behavior 
reverses, that is, for large $\phi$ there is a crossover were the diffusion
grows with $e_n$.
Theoretical evidence for an inverse relationship 
between diffusion and restitution coefficients
has recently been found in \cite{Zippelius}, as shown in their Fig.~(5). 
This is also noticeable in our Fig.~(\ref{trayectorias}), where one can observe 
how the particle excursions are larger for lower $e_n$.

The increase in diffusion for smaller $e_n$ is related to 
the appearance of density fluctuations that become stronger
as the dissipation increases. This weak clusterization becomes apparent as
a fall of the tails of the PDFs for large distances, as we will see later. 
Clusterization ends up freeing some
space in the system, allowing in this way faster diffusion. The same phenomenon is 
well known in colloid-polymer mixtures, where the depletion forces induced by 
the (small) polymers create strong density fluctuations in the colloidal phase, 
and increase its diffusion
\cite{Pham_Science296,Juarez_PRE77}.

\section{Granular Temperature and Velocity Distribution Functions}

In this section we perform a detailed analysis of the granular temperature $T_g$ and
of the Velocity Distribution Functions (VDFs) $P(v)$ obtained in the simulations. In 
Fig.~(\ref{Granular_Temperature}) we can observe the behavior of the in-plane $T_g$, 
defined as 
$T_g = T_x + T_y = \langle v_x^2 \rangle + \langle v_y^2 \rangle$, 
as a function of the packing fraction $\phi$, for all values of $\Gamma$ and 
$e_n$ considered. The following points deserve to be mentioned: first, for any given 
values of $\phi$ and $e_n$, $T_g$ increases together with $\Gamma$, as is intuitively 
expected, and experimentally observed \cite{Reis_PRE75}. Second, 
one can observe that for given values of $\phi$ and $\Gamma$,
$T_g$ decreases together with $e_n$, also as expected. 
Third, and more interesting, we find that for $e_n = 1$ the $T_g$ grows slowly 
with $\phi$, in agreement with the experimental results 
obtained in \cite{Reis_PRE75}. In that experiment $T_g$ grows 
monotonously until reaching a maximum for some  $\phi_{\mathrm{max}} \approx 0.5$, and 
then decays as $\phi$ keeps growing. We do not find this later decay of $T_g$, 
probably because we are using values of $\phi$ below $\phi_{\mathrm{max}}$. 

The coincidence found for 
this particular value of $e_n$ deserves some discussion: although the
authors of \cite{Reis_PRE75} do not report their working value 
of $e_n$, other references \cite{Urbach_PRL81,Kudrolli_PRL78} give $e_n \approx 0.9$ 
for the stainless steel spheres used there. 
As relates to the simulation, we remind the reader that even when $e_n = 1$ there is some 
dissipation in our system, since this value of the restitution 
coefficient is applied only 
for sphere-sphere and sphere-hemisphere collisions. Contacts with the upper 
plate (and, even if infrequent, with the lower plate),
are still inelastic, with $e_n = 0.878$. Besides, the existence of a nonzero 
value of $\mu$ gives us some energy losses via friction, for all collisions. 
One may therefore assume that there is, in our simulation, some effective $e_n$ a bit 
smaller than 1, 
and that the slow growth of the $T_g$ for diluted gases found 
in both experiment and simulation occurs for high values of $e_n$,
not necessarily $e_n = 1$.
This should be stressed since, for values of $e_n$ well bellow $1$, $T_g$ 
actually decays very slowly as $\phi$ increases, at least for the range of 
$\phi$ considered here. We are not aware of any experimental study of $T_g$ 
for highly dissipative materials, and so there is no verification of 
this curious behavior reversal for small restitution coefficients.

The VDFs for both components of the horizontal velocity have been obtained,
for all 
%velocity $P(v_i)$, where index $i$ makes reference to axis $x$ or $y$,  for the two 
values of $\Gamma$, $\phi$ and $e_n$ considered.
In Fig.~(\ref{colpDV}) we show a few of those curves, 
normalized by their characteristic velocities
$v_c=\sqrt{\left\langle v_i^2 \right\rangle} = \sqrt{T_i}$. 
In the same Figure we also show, as a reference, a unitary Gaussian. 
The first remarkable fact is that for all values of $\phi$, when
$e_n = 0.66$ and $1.0$, with $\Gamma = 3.5$, and also for all values of 
$e_n$ and $\Gamma = 7.2$, the VDFs present a strong peak at the center 
of the distribution ---that is, for low velocities---.
(Only some examples are shown, to avoid crowding the Figure.)
These peaks can be also found in Fig.~(8) of \cite{Reis_PRE75}, for 
small $\phi$. In that reference, they are attributed to the fact that
at these packing fractions the effect of interparticle collisions is minimal,
and so the general distribution is dominated by the VDF corresponding
to one isolated particle. The simulations show that
the main reason for the appearance of this peak is the friction of
the grains with the upper plate in the system. In particular, as can be 
easily visualized in a 1D system, rotational velocities acquired at the
collisions with the floor act as a braking factor upon contact with
the ceiling, increasing in this way the concentration of particles
with low horizontal velocities. In Fig.~(\ref{comparacion_mu_cero})
we show the normalized curves for the VDFs, for the case $\phi=0.20$, 
$\Gamma = 3.5$, and $e_n = 0.66$ and $1.0$, but taking the grain-ceiling
value of $\mu$ as $0.0$ in one case and $\mu = 0.25$ in the other. 
It is quite apparent from this Figure how, when there is no friction with
the ceiling, there are no low-velocity peaks.
However, it can be seen from the same Figure that the braking effect of the
ceiling is not the only factor involved in the non-Gaussian behavior of the 
VDFs; in particular, the line for $e_n = 0.66$ and $\mu_{\mathrm{ceiling}} = 0$
shows very clear deviations from the Gaussian. Even stronger deviations are found
for $e_n = 0.36$ and $\mu_{\mathrm{ceiling}} = 0$ (not shown in the Figure.) 
We would also like to mention that 
these peaks diminish as the packing fraction of the systems grows, as
observed in \cite{Reis_PRE75}.

To finish this section, let us show some analysis of the tails of the VDFs.
In Figs.~(\ref{Tails_1}) and~(\ref{Tails_3}) we plot 
$-\log(-\log(P(v/v_c)/P(0)))$ vs.  $\log(v/v_c)$ for $\Gamma = 3.5$ 
and $e_n = 0.36$, and for $\Gamma = 7.2$ and $e_n = 1.0$, as a function of $\phi$. 
For a perfect Gaussian centered at the origin, this plot should give a straight line
with slope $m = -2$.
It can be observed from both Figures how the behavior of the tails of the VDFs change slowly
with $\phi$, but quite strongly as $e_n$ is varied. We include in these 
graphs linear fits for the intermediate and final parts of the distribution.
Here it is important to notice that for $\Gamma = 3.5$ y $e_n = 0.36$, both 
the intermediate and final parts of the VDFs are close to a Gaussian, as
can be seen also, for $\phi = 0.15$,
in Fig.~(\ref{colpDV}). However, when $e_n = 1.0$, we find the intermediate
and final parts of the distribution to be clearly different. While the final part
is reasonably close to a Gaussian ($m_t \approx 2$) the intermediate part has an exponent 
larger than $-1$. This is due to the effect of the ceiling, since the deformation
of the VDFs ---that is, their deviation from a Gaussian--- 
in this region is related to the increase in the number of slow 
particles, which is in turn due to the friction with the upper plate. 
This general behavior can be corroborated in Fig.~(\ref{Tails_3})
($\Gamma = 7.2$, $e_n = 1.0$). Here we can see how 
the final part of the distribution remains lose to Gaussian, while one
gets an exponent $m_t \approx -0.5$, much larger than $-2$, for the 
intermediate part of the distribution. 

One then finds that the effect of the rough substrate is to return the form of the
VDFs to the Gaussian form, since the presence of the scatterers breaks translational 
symmetry and also forces a better horizontal energy interchange between the
particles. On the contrary, the effect of the friction with the ceiling
creates an increase in the distribution for slow particles, with a corresponding drop
for the distribution at intermediate velocities. 

\section{Pair distribution functions 
%and effective potentials
} 

The Pair Distribution Functions (PDFs) for a packing fraction $\phi = 0.35$ 
are shown in Fig.~(\ref{gder}), for some combinations of $\Gamma$ and $e_n$.
It can be observed how when $e_n = 1$, the PDFs do not depend on $\Gamma$,
and the structure observed is very similar to that of an equilibrium 
elastic hard sphere gas \cite{Urbach_PRE60}, indicating that the 
correlations that exist are mostly due to excluded volume effects. 
On the other hand, when the dissipation between particles increases, 
the correlations between particles begin to grow, an effect that is 
more notorious for low values of $\Gamma$. 

The lines shown in the graph have two intriguing characteristics:
first, we find a 
complete absence of any structure at $r/\sigma = \sqrt{3} = 1.73$,
where a peak appears in the close-packed limit.
Neither there is a peak around $r/\sigma = \sqrt{2} = 1.41$, which would 
signal some type of square clustering. It is therefore clear that the rough floor 
eliminates completely the possibility of formation of large crystalline
clusters. Second, a very peculiar secondary peak develops,
for small driving and high dissipation ($\Gamma = 3.5$ and $e_n = 0.36$),
at $r/\sigma \approx 2$. 
The possible reason we have identified for 
the origin of this peak is the formation, at this low values
of driving and restitution coefficient, of short-lived linear chains of grains
(see Fig.~(\ref{top_view_gg}) for an example). These linear chains may be induced by some
residual order in the substrate, an order that in turn is due to the fact that
it has to satisfy simultaneously a large density of scatterers and 
the no-overlap condition. It will be therefore interesting to see the evolution of
$g(r)$ in those experiments where an ordered substrate has been used 
\cite{Prevost_PRL89,Baxter_Nature2003};
although it is also clear from the graph that only for the smallest restitution 
coefficient used we get this peak.

Fore a more comprehensive view of the evolution of $g(r)$, 
in Fig.~(\ref{gder_gam3.5_1}) we show
the PDFs for $\Gamma = 3.5$, as a function of $e_n$ and $\phi$.
The very strong effect that $e_n$ has in these curves is the 
first thing to notice; for high dissipation the structure of
the main peak in $g(r)$ is almost independent of $\phi$, and 
little variation can be seen in the rest of the curve.
The large value of $g(r)$ at the main peak signals a tendency to
clustering. At the other extreme, the gas with $e_n = 1.0$
shows much more sensitivity to $\phi$. An unusual behavior can be seen for
$e_n = 1$ and small $\phi$: the largest value of 
$g(r)$ no longer falls at contact, and instead, a soft ``hill'' 
develops around $r/\sigma \approx 1.2$--$1.3$. This is again an effect of the
scatterers, that a this high value of $e_n$ intrude between the
grains, and, for small packing factors, reduce the probability
of contact. Again, there is no signal of any peak at $r/\sigma = \sqrt{3}$,
but we find instead that for the most dissipative system, some peak at 
$r/\sigma = 2$ begins to form. Visual inspection of the dynamics
in this parameter sector shows in effect the presence of short-lived
linear chains. For $\Gamma = 7.2$ (not shown) the results found are analogous:
there is an increase in the height of the main peak of $g(r)$ as $e_n$ is lowered,
although not as strong as for $\Gamma = 3.5$; and there is absolutely 
no peak at $r/\sigma = \sqrt{3}$. As for the behavior around $r/\sigma = 2$, 
now only some weak increase in $g(r)$ can be detected. A remarkable
result is the fact that for all $\phi$ used the different $g(r)$ obtained for 
$e_n = 1$ agree completely for both values of $\Gamma$.

\begin{figure} %\begin{indented} %\item[] 
\includegraphics[width=12.0cm,angle=270]{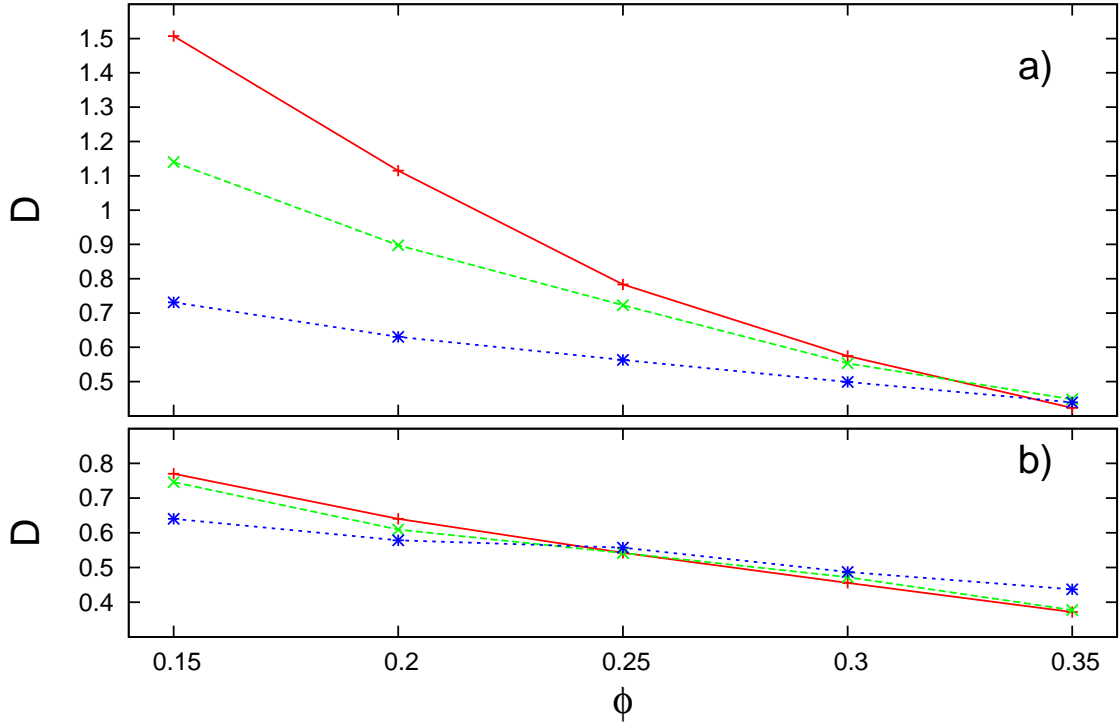} 

\caption{Diffusion Coefficients vs Filling Fraction for: (a) $\Gamma = 3.5$, 
and (b) $\Gamma = 7.2$. The values of $e_n$ are, at the left and starting from top:
$e_n = 0.36$ (red), $e_n = 0.66$ (green) and $e_n = 1$ (blue).
For the range of packing fractions covered, $D$
always decreases with $\phi$. The behavior with respect to $e_n$ is more complex: 
for small $\phi$, $D$ decays as $e_n$ grows. For large $\phi$ there seems to be a crossover
where $D$ starts growing with $e_n$.} 
\label{diff} 
\end{figure}

\begin{figure} %\begin{indented} %\item[] 
%\onefigure[width=5.5cm,angle=270]{figuras/GranularTemp_1.ps} 
\includegraphics[width=12.0cm,angle=270]{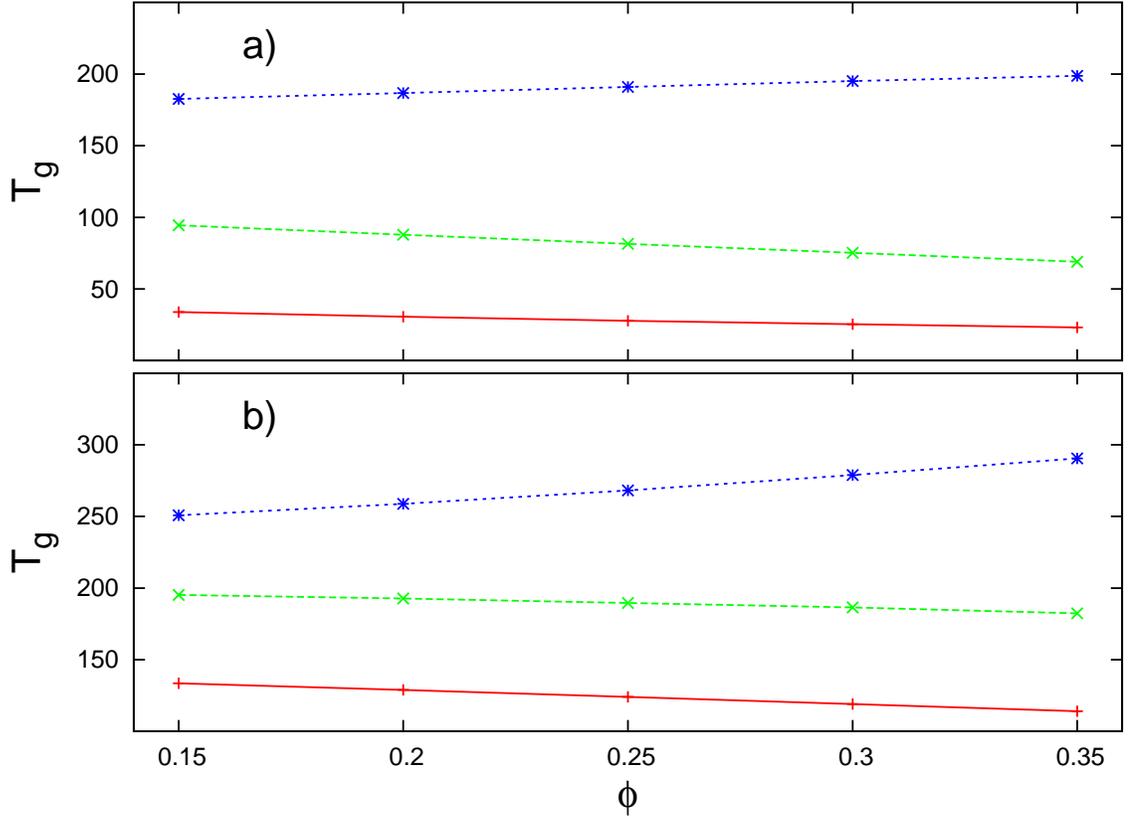} 

\caption{$T_g$ vs $\phi$ for both values of (a) $\Gamma = 3.5$ and (b) $\Gamma = 7.2$.
The values of $e_n$ are, for both graphs and starting from top:
$e_n = 1$ (blue), $e_n = 0.66$ (green), $e_n = 0.33$ (red). 
The dependence of $T_g$ on $\phi$ is much weaker than its dependence on $e_n$.
For both values of $\Gamma$ we find $T_g$ decaying slowly with $\phi$ at low values of
$e_n$, but increasing, agin slowly, with $\phi$ for $e_n = 1$.} 
\label{Granular_Temperature} 
\end{figure}

\begin{figure} %\begin{indented} %\item[] 
%\onefigure[width=5.0cm,angle=270]{figuras/Colp_D_vel.gamall.eall.nall.ps} 
\includegraphics[width=11.0cm,angle=270]{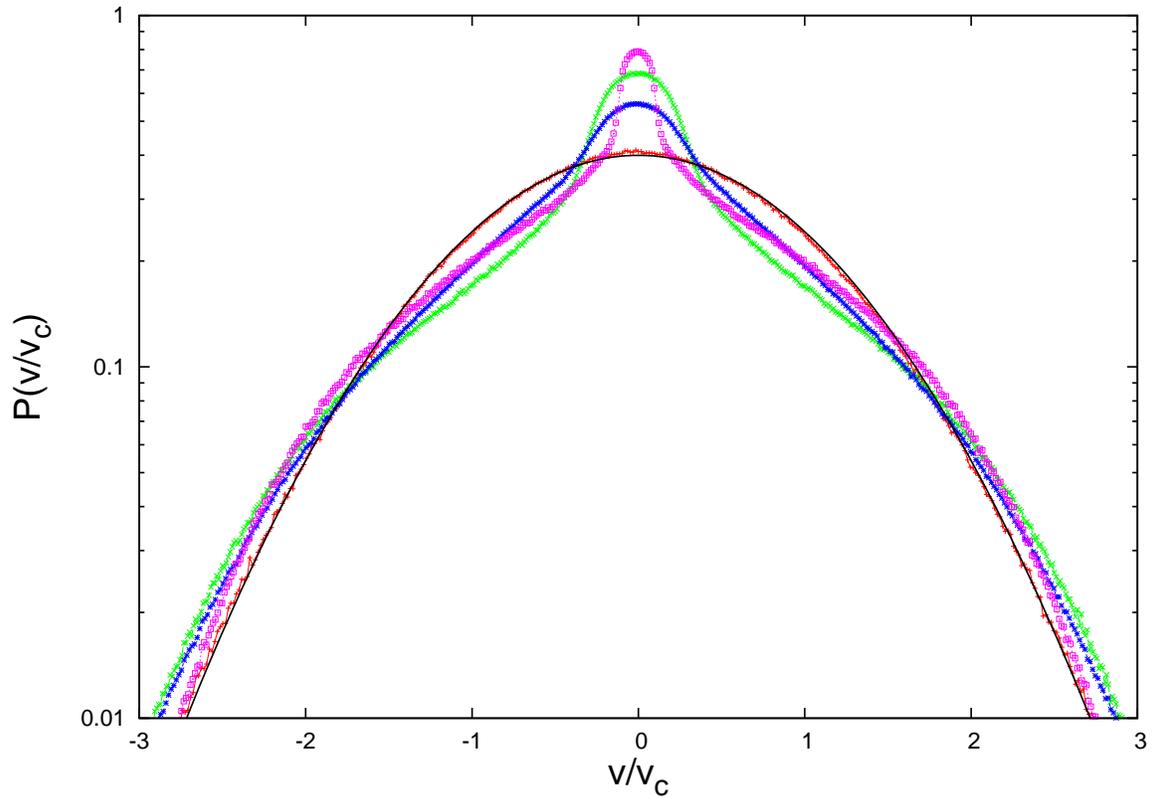} 

\caption{Velocity Distribution Functions (VDFs) for some values of the parameters. Here
the VDFs have been scaled by the RMS velocity. 
Looking at the center, and starting from
the top, the four thick lines correspond to: 
$\Gamma = 7.2$, $\phi = 0.15$ and $e_n = 0.36$ (magenta);
$\Gamma = 3.5$, $\phi = 0.15$ and $e_n = 1.0$ (green);
$\Gamma = 3.5$, $\phi = 0.35$ and $e_n = 1.0$ (blue); and 
$\Gamma = 3.5$, $\phi = 0.15$ and $e_n = 0.36$ (red).
The thin black line corresponds to a Gaussian distribution. 
For the last set of parameters the combination of high dissipation and small $\Gamma$
results in trajectories where the grains almost never touch the ceiling,
giving therefore a VDF very close to the Gaussian.} 
\label{colpDV} 
\end{figure}

\begin{figure} %\begin{indented} %\item[] 
%\onefigure[width=5.5cm,angle=270]{figuras/Comparation_mu.ps} 
\includegraphics[width=11.0cm,angle=270]{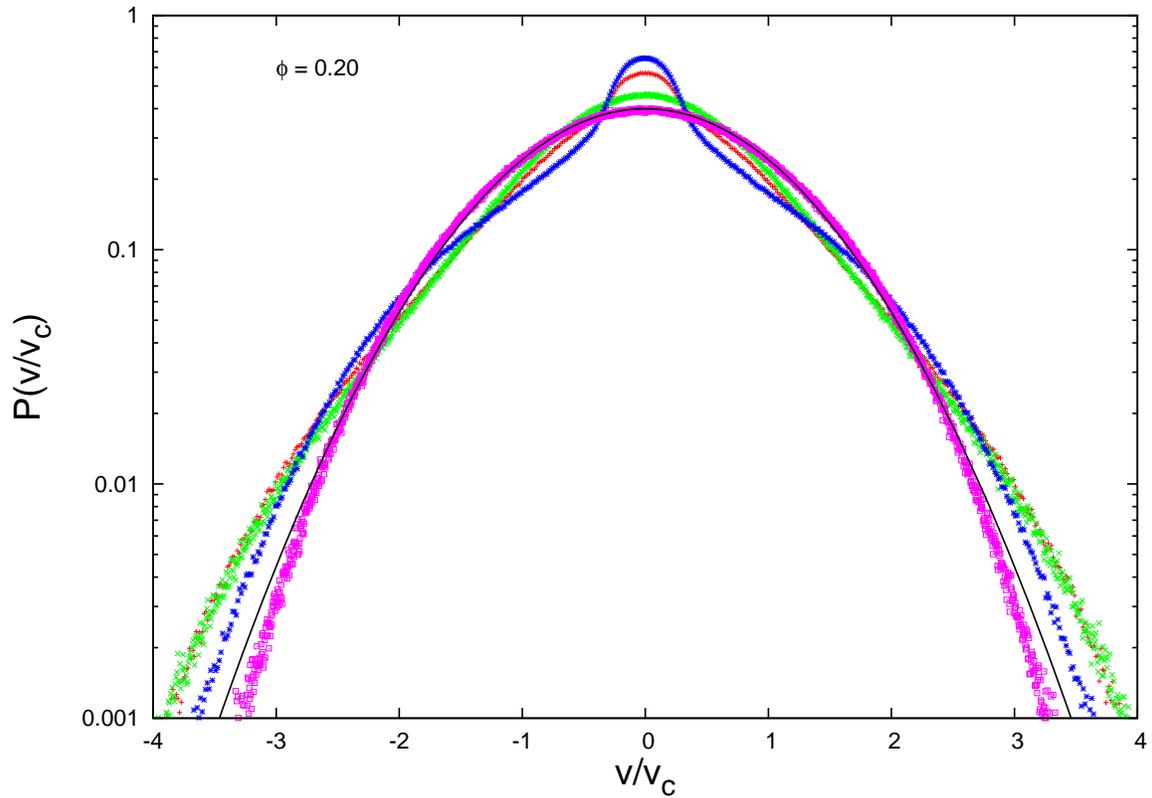} 

\caption{Some VDFs with and without friction with the ceiling.
For all lines we have used here $\Gamma = 3.5$ and $\phi = 0.2$.
Looking at the center, and starting from
the top, the four thick lines correspond to 
(a) $e_n = 1.0$ and $\mu = 0.25$;
(a) $e_n = 0.66$ and $\mu = 0.25$;
(a) $e_n = 0.66$ and $\mu = 0$; and
(a) $e_n = 1.0$ and $\mu = 0$.
The thin black line is again a Gaussian distribution.}
\label{comparacion_mu_cero} 
\end{figure}

\begin{figure} %\begin{indented} %\item[] 
%\onefigure[width=5.5cm,angle=270]{figuras/Tail.gam3.5.e0.5.nall.ps} 
\includegraphics[width=10.0cm,angle=270]{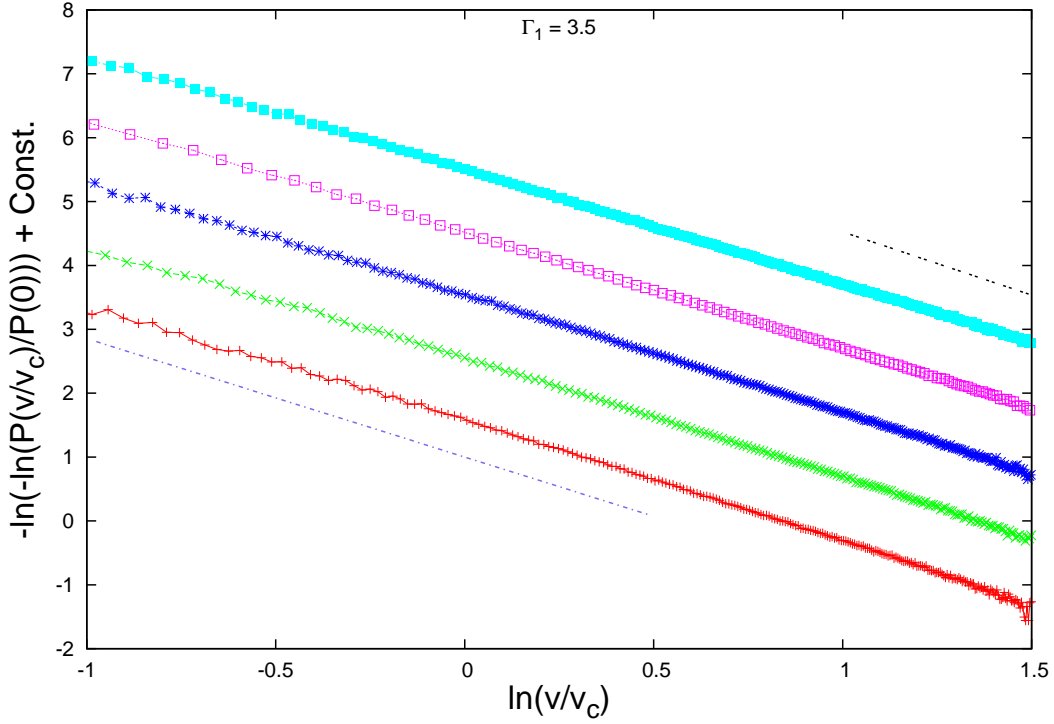} 

\caption{Intermediate and outer parts of the VDFs for $\Gamma = 3.5$ and $e = 0.36$, for
different values of $\phi$. 
Starting from top, the values of $\phi$ are $0.15$, $0.20$, $0.25$, $0.30$,
$0.35$. For clarity, lines have been shifted apart from each other.
Again, low $\Gamma$ and $e_n$ reduce contacts with
the ceiling, therefore generating VDFs very
close to Gaussians. The slope $m$ in the intermediate region is $m = -1.85$ (thin blue segment), 
and for the outer region (the tails) we get $m = -1.97$ (thin black segment).} 
\label{Tails_1} 
\end{figure}

\begin{figure} %\begin{indented} %\item[] 
%\onefigure[width=5.5cm,angle=270]{figuras/Tail.gam7.e0.5.nall.ps} 
\includegraphics[width=10.0cm,angle=270]{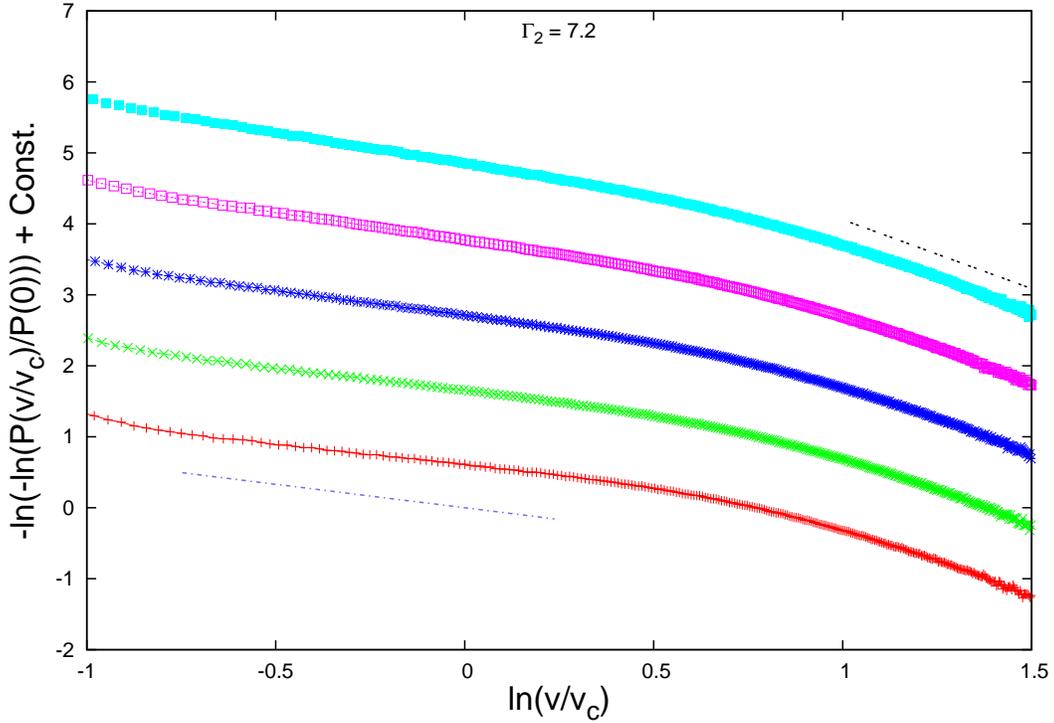} 

\caption{Intermediate and outer parts of the VDFs for $\Gamma = 7.2$ and $e = 1.00$, for
different values of $\phi$. The values of $\phi$ are as in the previous Figure,
and again the lines have been shifted apart from each other.
Here the high $\Gamma$ and $e_n$ imply strong bounces in the ceiling,
and therefore a very noticeable shift of the VDF towards low velocities. These VDFs are 
quite distorted, and the double-log vs log graph cannot be fitted well with 
straight lines. The best linear fits for the intermediate and outer regions 
give $m = -0.66$ (thin blue segment) and 
$m = -1.94$ (thin black segment), respectively.} 
\label{Tails_3} 
\end{figure}

\begin{figure} %\onefigure[width=5.5cm,angle=270]{figuras/RDF.gamall.eall.n588.ps} 
\includegraphics[width=10.0cm,angle=270]{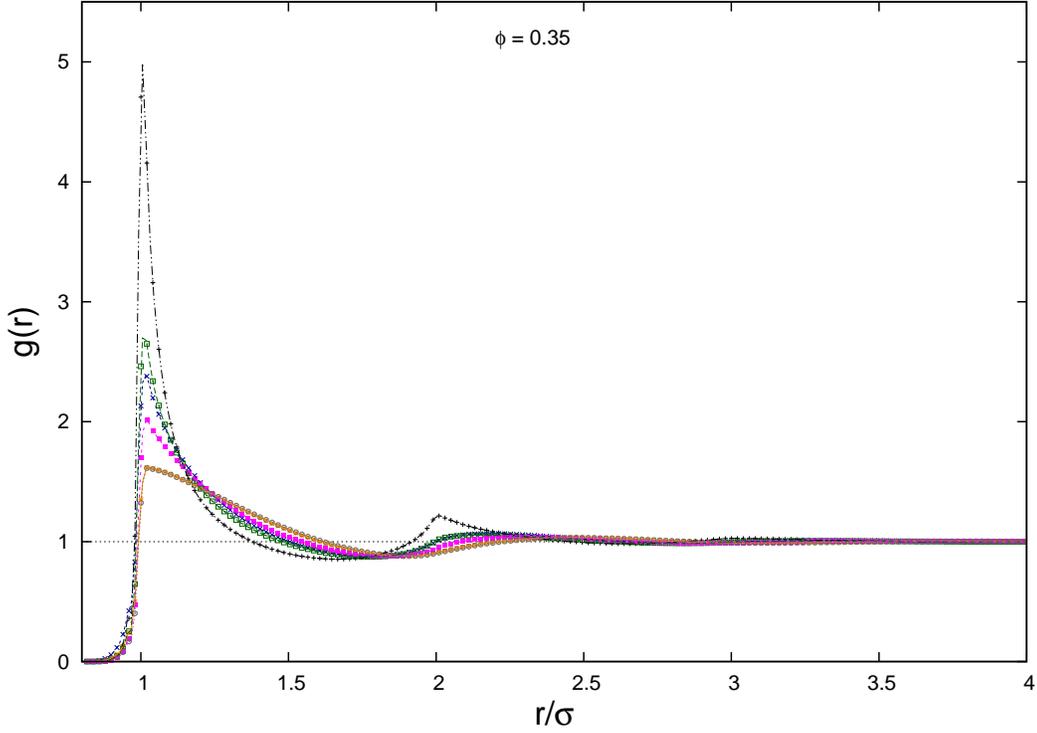} 

\caption{Pair distribution functions for $\phi = 0.35$ for some values of $\Gamma$ and $e_n$.
At the main peak, and starting from top, the changing parameters are:
(a) $\Gamma = 3.5$, $e_n = 0.36$;
(b) $\Gamma = 7.2$, $e_n = 0.36$;
(c) $\Gamma = 3.5$, $e_n = 0.66$;
(d) $\Gamma = 7.2$, $e_n = 0.66$;
(e) $\Gamma = 3.5$, $e_n = 1.0$ and
(f) $\Gamma = 7.2$, $e_n = 1.0$.
The last two lines fall on top of each other, showing how the PDFs for non-dissipative 
grains are independent of the acceleration. In general, smaller restitution coefficients
give rise to larger values of $g(r)$ at contact.} 
\label{gder} 
\end{figure}

\begin{figure} 
%\onefigure[width=5.5cm,angle=270]{figuras/RDF.gam3.5.eall.ngall_1.ps} 
\includegraphics[width=12.0cm,angle=270]{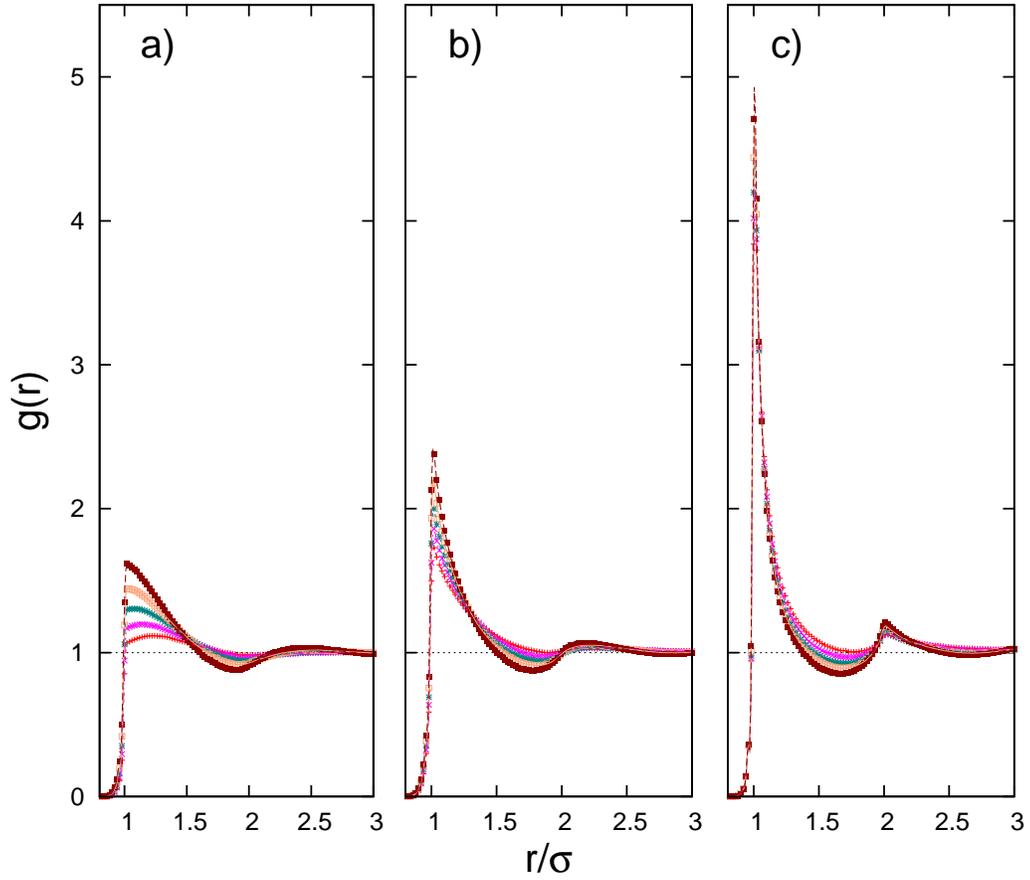} 

\caption{Pair distribution functions for $\Gamma_1 = 3.5$, for:
(a) $e_n = 1.0$,
(b) $e_n = 0.66$, and
(c) $e_n = 0.36$. 
In all cases, the $g(r)$ with the largest variation corresponds to $\phi = 0.35$,
and $g(r)$ becomes progressively flatter as $\phi$ decreases. Notice that in
(a), for $\phi < 0.30$, the largest peak is not located at contact, indicating
that the fixed scatterers have a weak separating effect over the grains.  
} 
\label{gder_gam3.5_1} 
\end{figure}

\begin{figure} 
%\onefigure[width=5.5cm,angle=270]{figuras/RDF.gam3.5.eall.ngall_2.ps} 
\includegraphics[width=10.0cm,angle=270]{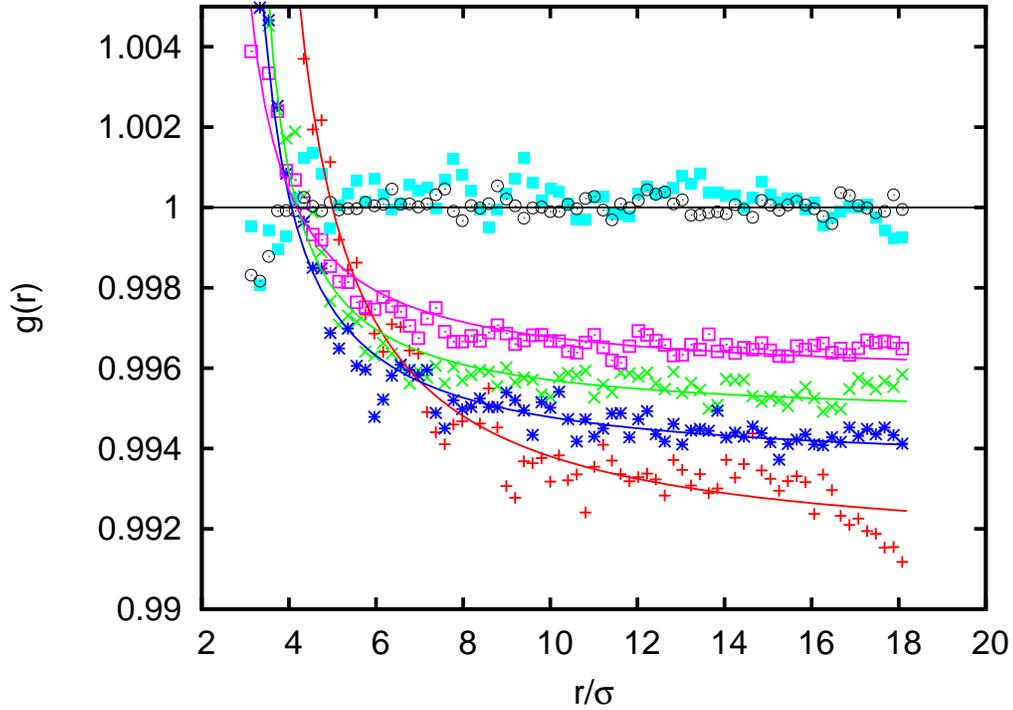} 
\caption{Zoom of the $g(r)$ for $\Gamma_1 = 3.5$. The decreasing of the $g(r)$
tails can be observed for different filling fractions with respect to the curve with $e = 1$.
The parameters are, starting from below:
(a) $e_n = 0.36$, $\phi = 0.15$ (red),
(b) $e_n = 0.66$, $\phi = 0.15$ (blue),
(c) $e_n = 0.36$, $\phi = 0.25$ (green), and
(d) $e_n = 0.66$, $\phi = 0.25$ (magenta).
The two sets of points running at $g(r) = 1$
correspond to $e_n = 1$, for $\phi = 0.15$ and $0.25$.
The lines are fits of the form $y(x) = A + B/(x-x_0)$,
and are provided only as guides to the eye; at this reduced
vertical scale, the large noise levels in the data make
a detailed evaluation of the parameters of the fit too uncertain as
to be useful.
}
\label{gder_closeup} 
\end{figure}

Finally, one of the most interesting points in the phenomenology
of this granular gas is shown in Fig.~(\ref{gder_closeup}), which displays
a close up to the PDFs for $\Gamma = 3.5$.
It can be observed here that for each filling fraction 
considered, $g(r)$ has its long-$r$ tail falling slightly below $1.0$ when 
we lower the restitution coefficient, something that does not happen
in the curves with $e_n = 1$. We would like to stress that the same 
normalization protocols were used for all lines. This behavior
implies a 
weak fractal behavior related to the non-unitary restitution 
coefficients in the system. More in detail, one can observe that the tails
in the PDFs show the largest drop below $1.0$ for $\phi = 0.15$ and
$e_n = 0.36$; in general, the PDFs approach $1.0$ from below as
$\phi$ increases or $e_n$ increase. As an effect of the clusterization
signaled by the falling of the PDFs bellow $1$, there is an increase
in the mean free path of those particles outside of the clusters
---that is, in the more diluted part of the gas---. This
gives rise to an increase of the average diffusion coefficient of the 
system, as was shown in Fig.~(\ref{diff}). 
%
%Finally, in Fig.~(\ref{gder_closeup})
For $\Gamma = 7.2$ one gets some appreciable decay in the
tails of $g(r)$ only for small $\phi$ and small $e_n$, and in general 
this effect is not so strong. This behavior 
again coincides with the increase in diffusion found for this 
value of $\Gamma$, as can be seen in Fig.~(\ref{diff}).

It is important to notice that the decay observed  in the tails of
the PFDs is due only to interparticle collisions, and independent of 
friction, either with the floor or the ceiling. This can be observed in 
Fig.~(\ref{gder_closeup}), where the fastest decay of the tails of the $g(r)$ 
appears for $\Gamma = 3.5$ y $e_n = 0.36$; for these values of the parameters the
VDFs are practically Gaussian, as can be seen in Fig.~(\ref{colpDV}), 
implying very little interaction with the ceiling. 

So, it is clear, looking at the results for diffusion and for the PDFs, 
that clustering increases the average diffusion, even if these clusters
are very unstable and their lifetimes are very short. The weakly fractal
behavior of the gas disappears when we increase any of the three controlling
parameters in the system: the packing fraction, the amplitude of vibration 
or the restitution coefficient.

\section{Conclusions and final comments}

We have performed numerical simulations of a vibrating quasi 2D granular 
gas, using a fully 3D algorithm with Coulomb friction proportional to normal 
forces, for a horizontal granular gas confined by a rough floor and a flat
ceiling. The system shows some of the characteristics already explored in 
experimental and other simulational work, as for instance the recovery 
of Gaussian VDFs at low $\Gamma$ over a rough substrate. The principal result 
of the simulation is how, even with a floor full of scatterers,
there is some level of clustering for the low range of $\Gamma$, $e_n$ and $\phi$
values. This clustering shows in the strong growth of the main peak of
$g(r)$, in the increase in diffusivity as $e_n$ is lowered, and very clearly in
the drop of $g(r)$ below 1 for for $r \gg \sigma$. 
All of the characteristics that indicate clustering 
coexist, however, with an almost complete recovery of Gaussian behavior in the VDFs 
for this part of the parameter space. For the other corner of this space
---that is, high $\Gamma$ and $\phi$ and $e_n = 1$--- all signals of clustering 
disappear, making it clear that the rough floor is only of minor importance 
with respect to this behavior. 

There are also some very punctual phenomena associated to this simulation: first,
it shows quite dramatically the distorting effect that the ceiling may have on the 
VDFs, by introducing a braking affect that results in a large increase in 
the central part of the distributions ---the very low velocity regime---. Experimental
results reported up to now show some hints of this problem, but there is not, 
to our knowledge, any systematic experimental exploration of this effect. 
Also, the presence of a peak in $g(r)$ around $r/\sigma = 2$ is pointing towards 
some complex effects of the substrate on the dynamics. Even if the configurations
of scatterers used in the simulation have a random origin, the simultaneous demands
large density and no overlap may have introduced some level of ordering, later
reflected is a larger-than-expected population of linear and almost-linear chains
in the granular gas. The appearance of undesired order in the substrate is not
surprising, since it is actually quite difficult to pack monodisperse disks 
densely without generating a crystal.

\acknowledgments

This work was supported by CONACyT (Mexico) through Grant No. 82975. J.\ A.\ P.-B.
receives support from a CONACyT Doctorate Fellowship.

\newpage

%=======================fig 01======================

%===================================================

%=======================fig 02====================== 
%===================================================

%=======================fig 03====================== %\onefigure{epl-template.eps} 
%===================================================

%=======================fig 04====================== %\onefigure{epl-template.eps} 
%===================================================

%=======================fig 05====================== %\onefigure{epl-template.eps} 
%===================================================

%=======================fig 6====================== %\onefigure{epl-template.eps} 
%===================================================

%=======================fig 7====================== %\onefigure{epl-template.eps} 
%===================================================

%=======================fig 8====================== %\onefigure{epl-template.eps} 
%===================================================

%=======================fig 9====================== %\onefigure{epl-template.eps} 
%===================================================

%=======================fig 10====================== %\onefigure{epl-template.eps} 
%===================================================

%=======================fig 11====================== %\onefigure{epl-template.eps} 
%===================================================

\end{document}